**Temperature oscillations in harmonic triangular lattice with random initial velocities**

V.A. Tsaplin[1], V.A. Kuzkin[1,2]

vtsaplin@yandex.ru, kuzkinva@gmail.com

[1]Institute for Problems in Mechanical Engineering, Bolshoy pr. V.O. 61, St. Petersburg, 199178, Russia

[2]Peter the Great Saint Petersburg Polytechnic University, Polytechnicheskaya str. 29, St. Petersburg, 195251, Russia

Abstract

Transition to thermal equilibrium in a uniformly heated two-dimensional harmonic triangular lattice with nearest neighbor interactions is investigated. Initial conditions, typical for molecular dynamics simulations, are considered. Initially, particles have uncorrelated random velocities, corresponding to initial kinetic temperature of the system, and zero displacements. These initial conditions can be realized by heating of the system by an ultrafast laser pulse. In this case, the kinetic temperature of the system oscillates. The oscillations are caused by the redistribution of energy between kinetic and potential parts. At large times, energies equilibrate and temperature tends to the equilibrium value equal to a half of the initial temperature. In our previous works, an integral exactly describing this transient thermal process has been derived. The integrand depends on the dispersion relation for the lattice. The integral contains large parameter, notably time. In the present work, we investigate large time behavior of the kinetic temperature. Simple asymptotic expression for deviation of temperature from the steady state value is derived. The expression contains three harmonics with different frequencies and amplitudes. Group velocities corresponding to these frequencies are equal to zero. Two frequencies are close and therefore beats of kinetic temperature are observed. Amplitude of deviation of temperature from the steady state value decreases inversely proportional to time. It is shown that the asymptotic formula has reasonable accuracy even at small times of order of one period of atomic vibrations.

**Keywords**: triangular lattice, harmonic crystal, kinetic temperature, beats, transition to equilibrium, nonequilibrium process.

## 1. Introduction

Description of nonequilibrium thermal processes in crystals is a long-standing problem in mechanics and physics of solids [1,2]. An example of such process is a transition of a system from nonequilibrium state towards thermal equilibrium. An initial nonequilibrium state can be caused, for example, by an ultra short laser pulse [3-7] or by propagation of shock waves [8,9]. In these cases, energy is unequally distributed among degrees of freedom and the kinetic temperature can demonstrate tensor properties [8,9]. After the end of nonequilibrium process, material relaxes towards thermal equilibrium. Kinetic and potential energies tend to equilibrium values.

Similar phenomenon is observed in the beginning of molecular dynamics simulations at finite temperatures [10]. In typical simulation, particles have random initial velocities and they are located at equilibrium positions. Then initially, total and kinetic energies of the system are equal. In the beginning of simulation, kinetic and potential energies oscillate in time and tend to the equilibrium values. The kinetic temperature, proportional to kinetic energy, also oscillates. In the present paper, we investigate this nonequilibrium process. Note that the process is independent on the initial distribution function for velocities. Behavior of the distribution function and its convergence to normal distribution is discusses in papers [11,12].

In harmonic crystals, transition to thermal equilibrium can be described analytically. In papers [13,14], the transition was investigated in harmonic one-dimensional chain. It has been shown that oscillations of kinetic temperature in a chain with random initial velocities and zero initial displacements is described by the Bessel function. In paper [15], similar problem was solved for one-dimensional chain with harmonic on-site potential. In papers [16,17] the results were generalized for the multidimensional case. In particular, an exact formula describing oscillations of kinetic temperature in harmonic triangular lattice has been derived.

In the present paper, we continue analysis started in papers [16,17]. In-plane motions of harmonic triangular lattice with nearest-neighbor interactions are considered. Initially particles have uncorrelated random velocities, corresponding to initial kinetic temperature of the system, and zero displacements. An integral exactly describing oscillation of temperature in this system has been derived in paper [16]. In the present paper, we consider long-time asymptotic behavior of this integral. It allows, in particular, to calculate characteristic frequencies of temperature oscillations and to explain physical meaning of these frequencies.

## 2. Oscillations of kinetic temperature: an exact solution

In the present section, we recall the exact expression describing oscillations of kinetic temperature in a uniformly heated harmonic triangular lattice [16].

Particles in the lattice are numbered by indices $n, k$ so that their radius vectors $\boldsymbol{r}_{n,k}$ have form [18,19]

$$\boldsymbol{r}_{n,k} = a(n\boldsymbol{e}_1 + k\boldsymbol{e}_2), \quad \boldsymbol{e}_3 = \boldsymbol{e}_1 + \boldsymbol{e}_2,$$

where $a$ is an equilibrium distance between the particles. Each particle interacts with six nearest neighbors via harmonic forces (see fig. 1). Anharmonicity of interactions is neglected. Anharmonic effects in triangular lattice are considered, for example, in papers [16, 20-22].

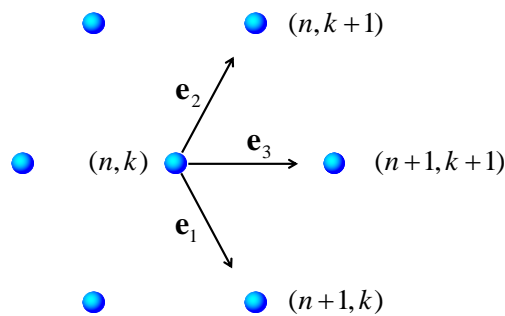

Figure 1. Particle with indices ($n,k$) and its nearest neighbors in triangular lattice.

Equations of motion of the particles have form

$$\ddot{u}_{n,k} = \omega_*^2 \Big( e_1 e_1 \cdot (u_{n+1,k} - 2u_{n,k} + u_{n-1,k}) + e_2 e_2 \cdot (u_{n,k+1} - 2u_{n,k} + u_{n,k-1}) + e_3 e_3 \cdot$$
$$(u_{n+1,k+1} - 2u_{n,k} + u_{n-1,k-1})\Big), \qquad (1)$$

where $\omega_* = \sqrt{C/m}$, $C$ is bond stiffness, $m$ is particle mass.

Initial velocities of the particles correspond to initial kinetic temperature of the system. Spatial distribution of temperature in the lattice is uniform. The velocities are independent random vectors with zero mean and equal variances. Initial displacements are equal to zero. In this case, initial kinetic energy is equal to the total energy of the system. In papers [16, 17] it is shown that the difference between kinetic and potential energies (Lagrangian) oscillates in time. It tends to zero as time tends to infinity, i.e. kinetic and potential energies equilibrate as predicted by the Virial theorem. The oscillations of the Lagrangian are described by the differential-difference equation similar to the equations of motion (1). This equation is solved using the discrete Fourier transform. The exact solution allows to calculate the deviation $\delta T$ of kinetic temperature from the steady state value:

$$\delta T = T - \frac{T_0}{2} = \frac{T_0}{4\pi^2}(I_1 + I_2), \quad I_k = \int_0^\pi \int_0^\pi \cos(\omega_k \tau)\, dp\, ds, \quad \tau = 2\sqrt{2}\omega_* t, \quad (2)$$

where $T_0$ is the initial kinetic temperature of the crystal; $\tau$ is dimensionless time; frequencies $\omega_1$ and $\omega_2$ are branches of the dispersion relation for the triangular lattice defined as non-negative solutions of equation

$$\omega_k^4 - 2B\omega_k^2 + C = 0, \quad B = \sin^2 p + \sin^2 s + \sin^2(p+s),$$
$$C = 3[\sin^2(p+s)(\sin^2 p + \sin^2 s) + \sin^2 p \sin^2 s]. \qquad (3)$$

In the following section, we focus on the large-time asymptotic behavior of kinetic temperature. Asymptotic analysis allows to calculate the decay rate and frequencies of temperature oscillations.

## 3. Large time behavior of kinetic temperature

In the present section, we derive simple formula describing large time behavior of the kinetic temperature using asymptotic methods [23,24].

Consider asymptotic behavior of integrals (2) at large time ($\tau \to \infty$). We reduce the integration domain using symmetry of functions $\omega_k(p, s)$. In particular, the functions satisfy the condition $\omega_k(p, s) = \omega_k(s, p)$. Using this and other symmetry conditions it can be shown that integrals over 12 minimal triangles in figure 2 are equal. Therefore integrals $I_k$ take the form

$$I_k = 12 \iint_{ABC} \cos(\omega_k(p, s)\tau)\, dp\, ds. \qquad (4)$$

We change variables $p, s$ to $\omega_1, \omega_2$ in integral (4). Then the triangle $ABC$ is mapped on the curvilinear triangle $O_1 O_2 O_3$ (see fig. 2).

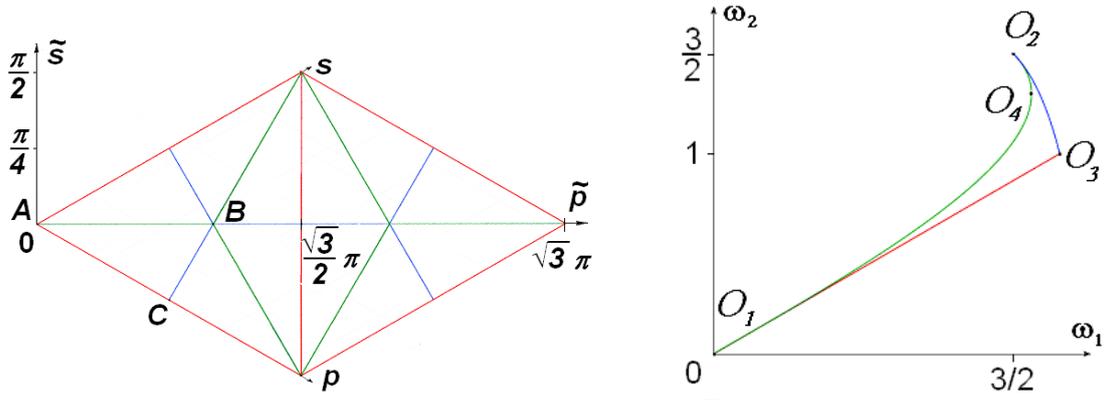

Figure 2. Integration domain in variables $\tilde{p} = \frac{\sqrt{3}}{2}(p+s)$, $\tilde{s} = \frac{1}{2}(s-p)$ (left). Integration region in the variables $\omega_1, \omega_2$ (right).

Using equations (3) and (A2), we obtain the following expressions for sides of the curvilinear triangle $O_1O_2O_3$:

$$
\begin{aligned}
AB \leftrightarrow O_1O_2: \quad & \omega_1 = \omega_2\sqrt{3 - \frac{8}{9}\omega_2^2}, \quad 0 \leq \omega_2 \leq \frac{3}{2} \\
BC \leftrightarrow O_2O_3: \quad & \omega_2 = \omega_1\sqrt{3 - \frac{8}{9}\omega_1^2}, \quad \frac{3}{2} \leq \omega_1 \leq \sqrt{3} \\
AC \leftrightarrow O_1O_3: \quad & \omega_2 = \frac{\omega_1}{\sqrt{3}}, \quad 0 \leq \omega_1 \leq \sqrt{3}
\end{aligned}
\quad (5)
$$

Changing variables in integral (4) yields

$$
I_k = 12 \iint_{O_1O_2O_3} J(\omega_1, \omega_2) \cos(\omega_k \tau)\, d\omega_1\, d\omega_2, \quad J(\omega_1, \omega_2) = \begin{vmatrix} \frac{\partial p}{\partial \omega_1} & \frac{\partial s}{\partial \omega_1} \\ \frac{\partial p}{\partial \omega_2} & \frac{\partial s}{\partial \omega_2} \end{vmatrix}, \quad (6)
$$

where $J$ is the Jacobian of the transformation. Note that components of the inverse Jacobian are equal to components of the group velocities.

After minor algebra in formula (6), we obtain the Fourier-type integrals:

$$
I_1 = \int_0^{\sqrt{3}} f_1(\omega_1) \cos(\omega_1 \tau)\, d\omega_1, \quad I_2 = \int_0^{\frac{3}{2}} f_2(\omega_2) \cos(\omega_2 \tau)\, d\omega_2,
$$

$$
f_1(\omega_1) = 12 \int_{\lambda_2(\omega_1)} J(\omega_1, \omega_2)\, d\omega_2, \quad f_2(\omega_2) = 12 \int_{\lambda_1(\omega_2)} J(\omega_1, \omega_2)\, d\omega_1,
$$
(7)

where $\lambda_2(\omega_1)$, $\lambda_1(\omega_2)$ are the integration regions determined by formulas (5).

In appendix B, it is shown that asymptotic behavior of integrals $I_k$ depends on the values of functions $f_k$ on the boundaries of the integration domain and on singularities of these functions. It can be shown that functions $f_k(\omega)$, defined by equations (7), are regular in the integration domain, except for points $O_1(0,0), O_2\left(\frac{3}{2},\frac{3}{2}\right), O_3(\sqrt{3},1), O_4\left(\frac{9}{4\sqrt{2}},\frac{3\sqrt{3}}{4}\right)$ in figure 2. Points $O_1$ and $O_2$ do not contribute to asymptotics of integrals $I_k$. The contribution is zero because $f_k(0) = 0$ at the beginning of the integration interval and $f_2(3/2) = 0$ at the end of the integration interval.

Functions $f_1, f_2$ have logarithmic singularities at the points $9/(4\sqrt{2})$ and $1$ respectively, i.e. in the vicinity of these points $f_1 \sim L_1 \ln|\omega_1 - 9/(4\sqrt{2})|$ and $f_2 \sim L_2 \ln|\omega_2 - 1|$. Contribution of these points to the asymptotics is proportional to coefficients $L_1, L_2$ calculated in appendixes C and D. Another contribution to asymptotics of the integral $I_1$ is from the value of function $f_1$ at the boundary point $\omega = \sqrt{3}$. This value is calculated in appendix C. Therefore asymptotic behavior is determined by three special points. Summing contributions of all special points yields:

$$I_1 \approx \frac{4\sqrt{3}\pi}{\tau}\left[2\sqrt{\frac{2}{7}}\cos\left(\frac{9}{4\sqrt{2}}\tau\right) + \sin(\sqrt{3}\tau)\right], \quad I_2 \approx \frac{12\pi}{\sqrt{7}}\frac{\cos\tau}{\tau}. \quad (8)$$

Substituting formula (8) into formula (2) we obtain the asymptotic expression describing large time behavior of the kinetic temperature:

$$\delta T \approx \frac{T_0}{2\sqrt{2}\pi\omega_* t}\left[2\sqrt{\frac{6}{7}}\cos\left(\frac{9}{2}\omega_* t\right) + \sqrt{3}\sin(2\sqrt{6}\omega_* t) + \frac{3}{\sqrt{7}}\cos(2\sqrt{2}\omega_* t)\right]. \quad (9)$$

Formula (9) describes deviation of temperature from the steady-state value. It shows that the deviation decays inversely proportional to time. Oscillations of temperature have three main frequencies

$$\Omega_1 = \frac{9}{2}\omega_*, \quad \Omega_2 = 2\sqrt{6}\omega_*, \quad \Omega_3 = 2\sqrt{2}\omega_*. \quad (10)$$

These frequencies correspond to three special points of integrals $I_k$. Note that frequencies $\Omega_1$ and $\Omega_2$ are close. Therefore beats of kinetic temperature are observed.

It can be shown using formulas (3) that group velocities corresponding to frequencies (10) are equal to zero:

$$\left.\frac{\partial \omega_{1,2}}{\partial p}\right|_{O_3} = \left.\frac{\partial \omega_{1,2}}{\partial s}\right|_{O_3} = 0$$

$$\left.\frac{\partial \omega_1}{\partial p}\right|_{O_4} = \left.\frac{\partial \omega_1}{\partial s}\right|_{O_4} = 0$$

Note that the same fact is observed in one-dimensional chain with harmonic on-site potential [15].

Comparison of the exact expression (7) for $\delta T$ with asymptotic formula (9) is shown in figure 3. For $\omega_* t/(2\pi) > 2.5$ the difference between exact and asymptotic formulas is less than 6%.

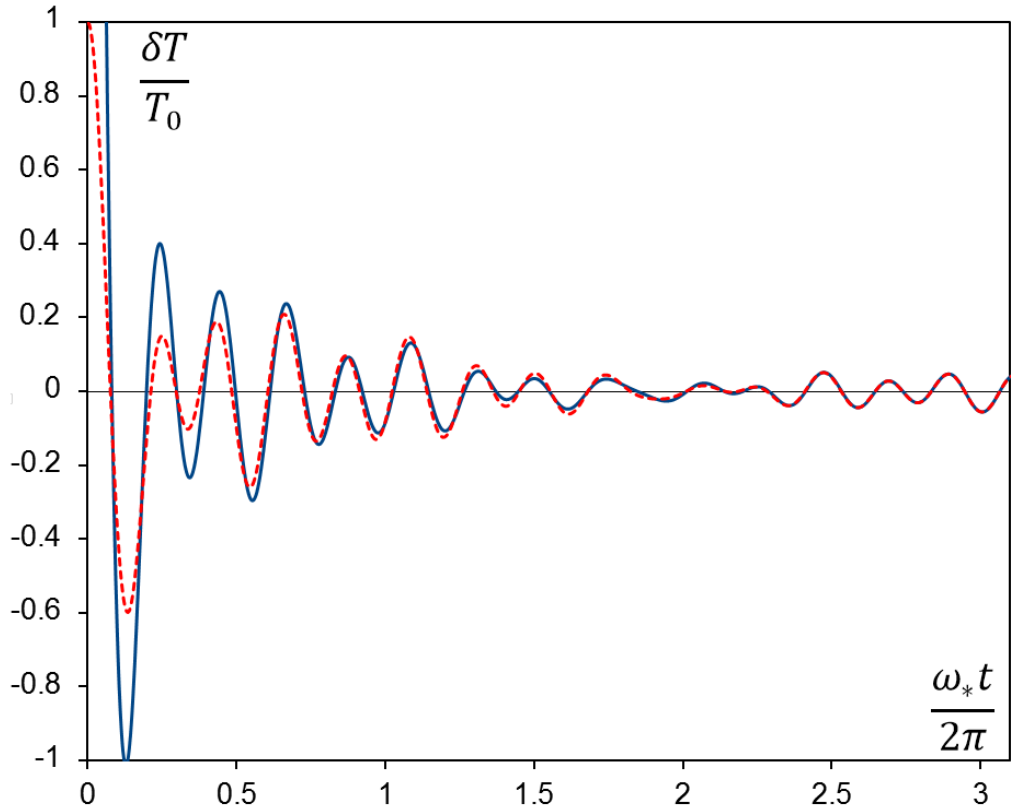

Figure 3. Oscillations of kinetic temperature in triangular lattice with random initial velocities and zero initial displacements. Solid line — asymptotic formula (9); dashed line — exact formula (2).

Thus formula (9) describes oscillations of kinetic temperature even at relatively small times $\omega_* t/(2\pi) \sim 1$.

## 5. Conclusions

Oscillations of kinetic temperature in harmonic triangular lattice with random initial velocities and zero initial displacements were considered. Simple asymptotic formula describing large time behavior of these oscillations was derived. Comparison with an exact solution shows that the asymptotic formula has reasonable accuracy even at small times of order of one period of atomic vibrations. It is shown that deviation of kinetic temperature from the steady state value is represented as a sum of three harmonics with different frequencies and amplitudes. The amplitudes are inversely proportional to time. Group velocities corresponding to frequencies in asymptotics are equal to zero. The same effect has previously been reported for one-dimensional chain with harmonic on-site potential [15]. Therefore we assume that in harmonic crystals frequencies of oscillations of kinetic temperature correspond to zero group velocities. However this hypothesis is awaiting a rigorous proof.

**Acknowledgements**

This work was supported by the Russian Science Foundation (RSCF grant № 17-71-10213).

**Appendix A. Change of integration variables**

Consider change of integration variables $p, s$ to $\omega_1, \omega_2$ in integrals (2). Variables $\omega_1, \omega_2$ satisfy the equation:

$$\omega_k^4 - 2B\omega_k^2 + C = 0. \qquad (A1)$$

Relations between $B, C$ and $p, s$ are given by formulas (3). On the other hand, according to Vieta's theorem and equation (A1)

$$B = \frac{\omega_1^2 + \omega_2^2}{2}, \quad C = \omega_1^2 \omega_2^2. \qquad (A2)$$

Using formulas (3), (A2) we make two consecutive changes of variables $p, s$ to $B, C$ and $B, C$ to $\omega_1, \omega_2$. The resulting Jacobian $J$ is represented in the form:

$$J(\omega_1, \omega_2) = |J_1|^{-1} J_2,$$

$$J_1 = \begin{vmatrix} \dfrac{\partial B}{\partial p} & \dfrac{\partial B}{\partial s} \\ \dfrac{\partial C}{\partial p} & \dfrac{\partial C}{\partial s} \end{vmatrix} = 12 \sin p \sin s \sin(p+s) \sin(p-s) \sin(p+2s) \sin(2p+s),$$

(A3)

$$J_2(\omega_1, \omega_2) = \begin{vmatrix} \dfrac{\partial B}{\partial \omega_1} & \dfrac{\partial B}{\partial \omega_2} \\ \dfrac{\partial C}{\partial \omega_1} & \dfrac{\partial C}{\partial \omega_2} \end{vmatrix} = 2(\omega_1^2 + \omega_2^2)\omega_1 \omega_2.$$

Note that the Jacobian $J$ is finite and it is not equal to zero inside the integration domain in formula (2). Therefore there is a unique correspondence between $p, s$ and $\omega_1, \omega_2$.

## Appendix B. Asymptotics of Fourier-type integrals

In the present appendix, we consider asymptotic behavior of the Fourier-type integral

$$I = \int_0^b f(\omega) \cos(\omega \tau) \, d\omega \qquad (B1)$$

for large values of parameter $\tau$. Assume that $f(\omega)$ is a polynomial. Then integration by parts in formula (B1) yields the following asymptotics:

$$I = f(b) \frac{\sin b\tau}{\tau} + O\left(\frac{1}{\tau^2}\right), \qquad \tau \to \infty. \qquad (B2)$$

Formula (B2) is valid for polynomial of any degree. Therefore it is also valid for function $f(\omega)$ being regular on the interval $[0, b]$.

Formula (B2) is generalized for functions having $N$ discontinuities on the integration interval. Assume that functions $f$ has $N$ finite jumps $G_i$ at points $\widetilde{\omega}_i \in (0, b)$:

$$G_i = \lim_{\varepsilon \to +0} \big(f(\widetilde{\omega}_i + \varepsilon) - f(\widetilde{\omega}_i - \varepsilon)\big). \qquad (B3)$$

Then separating the interval $(0, b)$ into $N + 1$ intervals, where $f$ is regular, and using formula (B2) yields:

$$I \approx f(b) \frac{\sin b\tau}{\tau} - \frac{1}{\tau} \sum_{i=1}^{N} G_i \sin \widetilde{\omega}_i \tau. \qquad (B4)$$

Formula (B4) shows that asymptotic behavior of integral (B1) depends on discontinuities of function $f$ and its value $f(b)$ at the boundary point.

Consider logarithmic singularity of function $f$ at the point $\omega = \omega'$, $(0 \le \omega' < b)$. In other words, in the vicinity of this point the function has form

$$f(\omega) \approx L \ln|\omega - \omega'|. \qquad (B5)$$

Substituting this expression into the integral (B1) yields:

$$I \approx \begin{cases} \dfrac{L}{\tau}(-\pi\cos\omega'\tau + \ln|b-\omega'|\sin b\tau), & \omega' > 0. \\ \dfrac{L}{\tau}\left(-\dfrac{\pi}{2} + \ln|b|\sin b\tau\right), & \omega' = 0. \end{cases} \quad (B6)$$

The contribution of the logarithmic singular point to asymptotics of the integral (B1) is given by the first terms in these expressions. The second terms is the contribution of the end of the integration interval $\omega = b$.

Generalizing formula (B4) for function $f(\omega)$ having $N$ discontinuities (each of them is jump or logarithmic singularity) at points $\tilde{\omega}_i \in [0, b)$, $i = 1 \ldots N$, yields:

$$\int_0^b f(\omega)\cos(\omega\tau)\,d\omega \approx f(b)\frac{\sin b\tau}{\tau} - \frac{1}{\tau}\sum_{i=1}^{N}(G_i\sin\tilde{\omega}_i\tau + \pi L_i\cos\tilde{\omega}_i\tau). \quad (B7)$$

Coefficients $G_i, L_i$ are defined by formulas (B3), (B5) respectively. If there is no logarithmic singularity at $\tilde{\omega}_i$, then $L_i = 0$. If $\tilde{\omega}_i = 0$, then corresponding coefficient $L_i$ in formula (B7) should be replaced by $L_i/2$ (see formula (B6)).

## Appendix C. Asymptotic behavior of functions $f_k$ in the vicinity of the point $O_3$

In the present appendix, we derive asymptotic expressions for functions $f_1(\omega_1), f_2(\omega_2)$, defined by formula (7), in the vicinity of the point $O_3$ ($\omega_1 = \sqrt{3}, \omega_2 = 1$). According to (B7), this point contributes to asymptotics of the integral (7).

We introduce new variables

$$\omega_1' = \omega_1 - \sqrt{3}, \quad \omega_2' = \omega_2 - 1. \quad (C1)$$

Then formula (7) for $f_2(\omega_2)$ reads

$$f_2(\omega_2) = 12\left(\int_{\text{bottom of }\lambda_1}^{\varepsilon} J'(\omega_1', \omega_2')\,d\omega_1' + \int_{\varepsilon}^{\text{top of }\lambda_1} J'(\omega_1', \omega_2')\,d\omega_1'\right),$$

$$J'(\omega_1', \omega_2') = J(\omega_1, \omega_2), \quad (C2)$$

where $\varepsilon$ is a small parameter. In the vicinity of the point $\omega_2 = 1$, the first term in formula (C2) is a regular function. Then discontinuity of the function $f_2(\omega_2)$ at this point is equal to discontinuity of the second term in formula (C2). Therefore it is sufficient to consider function $J'(\omega_1', \omega_2')$ in a small vicinity of the point $\omega_1' = 0, \omega_2' = 0$. Formulas (A3), (A2), (3) yield the following expression for small $\omega_1', \omega_2'$:

$$J'(\omega_1', \omega_2') \approx \frac{1}{\sqrt{(\omega_1' - \sqrt{3}\,\omega_2')(7\omega_1' + \sqrt{3}\,\omega_2')}}. \quad (C3)$$

We approximate the integration region $\lambda_1$ in the vicinity of the point $\omega_1' = 0, \omega_2' = 0$. Linearization of formulas (5) shows that the top of $\lambda_1$ in (C2) is calculated as $\omega_1' = \sqrt{3}\,\omega_2'$ for $\omega_2' < 0$ and $\omega_1' = -\sqrt{3}\,\omega_2'/7$ for $\omega_2' > 0$. Substitution of these expressions and formula (C3) into formula (C2) yields the following asymptotics for discontinuity of function $f_2(\omega_2)$ at $\omega_2 \to 1$:

$$f_2 \approx \frac{12}{\sqrt{7}} \ln|\omega_2 - 1|. \quad (C4)$$

Analogously, we calculate the value of function $f_1(\omega_1)$ in the vicinity of the point $\omega_1 = \sqrt{3}$. Integrating $J'(\omega_1', \omega_2')$ with respect to $\omega_2'$ in a small vicinity of the point $\omega_1' = 0, \omega_2' = 0$ yields

$$\lim_{\omega_1 \to \sqrt{3}-0} f_1(\omega_1) = 4\sqrt{3}\pi. \quad (C5)$$

Since the point $\omega_1 = \sqrt{3}$ is the end of integration interval, the value $f_1(\sqrt{3})$ contributes to asymptotics (see formula (B7)).

### Appendix D. Asymptotic behavior of functions $f_k$ in the vicinity of the point $O_4$

In the present appendix, we show that in the vicinity of the point $O_4$ ($\omega_1 = \frac{9}{4\sqrt{2}}$) function $f_1(\omega_1)$ has the following asymptotics

$$f_1 \approx L_1 \ln\left|\omega_1 - \frac{9}{4\sqrt{2}}\right|.$$

Parameter $L_1$ is calculated below. The following new variables are introduced

$$\omega_1' = \omega_1 - \frac{9}{4\sqrt{2}}, \quad \omega_2' = \omega_2 - \frac{3\sqrt{3}}{4}.$$

Then function $f_1$ takes the form

$$f_1(\omega_1) = 12 \int_{\lambda_2(\omega_1)} J'(\omega_1', \omega_2')\, d\omega_2', \quad J'(\omega_1', \omega_2') = J(\omega_1, \omega_2). \quad (D1)$$

To calculate the asymptotic of the discontinuity of function $f_1$ it is sufficient to consider integration in a small vicinity of the point $\omega_1' = 0, \omega_2' = 0$ (see appendix C). For small $\omega_1'$, $\omega_2'$ formulas (A3), (A2), (3) yield the following expression for $J'$:

$$J'(\omega_1', \omega_2') \approx \frac{2\sqrt{2}}{\sqrt{21\left(\omega_1'^2 + \frac{3\omega_2'}{4\sqrt{2}}\right)}}. \quad (D2)$$

Using formulas (5), we approximate the integration region $\lambda_2$ in the vicinity of the point $\omega_1' = 0, \omega_2' = 0$ by parabola $\omega_2' = -\frac{4\sqrt{2}}{3}\omega_1'^2$. Substituting this approximation and formula (D2) into integral (D1) yields the following asymptotics for discontinuity of function $f_1(\omega_1)$ at $\omega_1 \to \frac{9}{4\sqrt{2}}$:

$$f_1 \approx -8\sqrt{\frac{6}{7}} \ln\left|\omega_1 - \frac{9}{4\sqrt{2}}\right|. \quad (D3)$$

Using the same approach, we show that function $f_2(\omega_2)$ has no discontinuities at the point $\omega_2 = 3\sqrt{3}/4$.